%% file: main.tex
\documentclass{achemso}

\setkeys{acs}{
  articletitle = true,
  etalmode = truncate,
  maxauthors = 100
}
\SectionNumbersOn

\usepackage[utf8]{inputenc}
\usepackage{achemso}
\usepackage{amsmath}
\usepackage{multirow}
\usepackage{adjustbox}
\usepackage{color}
\usepackage{textcomp}

\newcommand{\denovo}{\textit{de novo }}

\newcommand{\newtilde}{\raisebox{0.5ex}{\texttildelow}}

\title{GuacaMol: Benchmarking Models for De Novo Molecular Design}

\author{Nathan Brown, Marco Fiscato$^*$, \protect\\ Marwin H.S. Segler$^*$, Alain~C.~Vaucher}
\date{November 2018}
\email{{marco.fiscato, marwin.segler, alain.vaucher}@benevolent.ai}
\affiliation{BenevolentAI, London, UK}

\begin{document}

\maketitle

\begin{abstract}
\textit{De novo} design seeks to generate molecules with required property profiles by virtual design-make-test cycles. With the emergence of deep learning and neural generative models in many application areas, models for molecular design based on neural networks appeared recently and show promising results. 
However, the new models have not been profiled on consistent tasks, and comparative studies to well-established algorithms have only seldom been performed. 
To standardize the assessment of both classical and neural models for \denovo molecular design, we propose an evaluation framework, GuacaMol, based on a suite of standardized benchmarks. The benchmark tasks encompass measuring the fidelity of the models to reproduce the property distribution of the training sets, the ability to generate novel molecules, the exploration and exploitation of chemical space, and a variety of single and multi-objective optimization tasks. The benchmarking open-source Python code, and a leaderboard can be found on \url{https://benevolent.ai/guacamol}.
\end{abstract}

\section{Introduction}
\textit{De novo} molecular design is a computational technique to generate novel compounds with desirable property profiles from scratch.\cite{nicolaou2013multi,schneider2013novo} It complements virtual screening, where large virtual compound libraries are pre-generated, stored, and then subsequently ranked on demand. Virtual screening is particularly useful when the virtual compounds are readily available (commercially, or in in-house screening collections), or easy to synthesize.\cite{irwin2005zinc, van2019virtual} 
Datasets sized on the order of $10^{13}$ can be routinely screened under current computational constraints.\cite{walters2018virtual} However, this is only a tiny fraction of drug-like chemical space, which has an estimated size anywhere between $10^{24}$ and $10^{60}$ possible structures.\cite{walters2018virtual,ruddigkeit2012enumeration} This number might be even larger considering new modalities such as PROTACs.\cite{valeur2017new} 
Efficient approaches to query larger combinatorial spaces without full enumeration via similarity to given structures have been reported.\cite{rarey2001similarity,parn2007exploring,lauck2013coping} 
However, it is not straightforward to perform more complex, multi-objective queries without enumerating substantial parts of the space.

In contrast, in \denovo design, only a relatively small number of molecules is explicitly generated, and chemical space is explored via search or optimization procedures. Thus, by focusing on the most relevant areas of chemical space for the current multi-objective query, \denovo design algorithms can in principle explore the full chemical space, given any ranking or scoring function.

In addition to established discrete models relying on molecular graphs,\cite{zaliani2009second,ertl2012iade,hartenfeller2011enabling} models for \denovo design based on deep neural networks have been proposed in recent years, and have shown promising results.\cite{segler_generating_2018,gomez-bombarelli_automatic_2018} 
Unfortunately, validation methods for these generative models have not been consistent. 
Comparative studies to established \denovo design models have not yet been performed. 
Furthermore, recent property optimization studies often focused on easily optimizable properties, such as drug-likeness or partition coefficients.\cite{you_graph_2018, liu_constrained_2018, gomez-bombarelli_automatic_2018} This makes it difficult to understand the strengths and weaknesses of the different models, to assess which models should be used in practice, and how they can be improved and extended further.

In other fields of machine learning, for example computer vision and natural language processing, standardized benchmarks have triggered rapid progress.\cite{deng2009imagenet, mayr2016deeptox} 
Similarly, we believe that the field of \denovo molecular design can benefit from the introduction of standardized benchmarks. They allow for a straightforward survey and comparison of existing models, and provide insight into the strengths and weaknesses of models, which is valuable information to improve current approaches. 

In this work, we introduce GuacaMol, a framework for benchmarking models for \denovo molecular design.
We define a suite of benchmarks and implement it as a Python package designed to allow researchers to assess their models easily. 
We also provide implementations and results for a series of baseline models on \url{https://benevolent.ai/guacamol}.

\section{Models for De Novo Molecular Design}

\textit{De novo} design approaches require three components: 1) molecule generation, 2) a way to score the molecules, and 3) a way to optimize or search for better molecules with respect to the scoring function.\cite{hartenfeller2011enabling} 
For scoring, any function that maps molecules to a real valued score can be used, for example quantitative structure-activity relationships (QSAR) and quantitative structure-property relationships (QSPR) models,  structure-based models for affinity prediction, heuristics for the calculation of physicochemical properties, and combinations thereof for multi-objective tasks.

At the center of molecule generation and optimization strategies lies the choice of the molecular representation, which determines the potential range of exploration of chemical space.
Molecules can be constructed atom-by-atom or by combination of fragments.\cite{hartenfeller2011enabling} They can be grown in the context of a binding pocket,\cite{bohm1992computer,gillet1994sprout,wang2000ligbuilder,nicolaou2009novo, chevillard2018binding} which allows for structure-based scoring, or be constructed and scored using entirely ligand-based methods. 
Choosing a very general representation of molecules for construction, such as atom-by-atom and bond-by-bond building, allows for potentially exploring all of chemical space. However, without any constraints or scoring defining what sensible molecules look like, such representations can lead to molecules which are very difficult to synthesize or potentially unstable, particularly in combination with powerful global optimizers such as genetic algorithms (GAs).\cite{brown2004graph,hartenfeller2011enabling} 
To alleviate this problem, different approaches were introduced to generate molecules using readily available building blocks and robust reaction schemes (e.g. amide couplings), or fragments derived from known molecules.\cite{chevillard2015scubidoo,hartenfeller2011enabling, degen2008art} This, however, may drastically limit the potentially explorable space. 
This compromise between realistic molecules and large explorable molecular space is a major dilemma of \denovo design. 

A seemingly obvious solution is to add a contribution penalizing unrealistic molecules to the scoring function. 
Unfortunately, it is not trivial to encode what medicinal chemists would consider as molecules that are acceptable to take forward to synthesis.\cite{lajiness2004assessment, segler2018planning} Instead of defining what good molecules look like as rules, or in the form of a scoring function, machine learning approaches represent an attractive alternative, because they are potentially able to learn how reasonable molecules look from data.\cite{schmidhuber2015deep,chen_rise_2018}  

Machine learning has been used in chemoinformatics for at least 50 years to score molecules (QSAR/QSPR applications), that is to predict properties given a structure.\cite{hansch1964p}
The inverse QSAR problem of predicting structures given target properties has received less attention.\cite{baskin1989methodology} 
However, recent advances in algorithms and hardware have made machine learning models and particularly neural networks, which directly output molecular structures tractable.\cite{schmidhuber2015deep}

Two early papers on generative models for molecules, first published as preprints, employed the simplified molecular-input line-entry system (SMILES) representation, which allows for the representation of molecular graphs as a string, with parentheses and numbers indicating branching points and ring closures.\cite{weininger1988smiles}
Gómez-Bombarelli et al.~proposed to use variational auto-encoders (VAEs), which encode the SMILES representation of a molecule into a latent, continuous representation in real space, and back.\cite{gomez-bombarelli_automatic_2018}
Optimization can then be performed by gradient descent or Bayesian optimization in real space. Segler et al.~introduced the notion of transfer and reinforcement learning for molecule generation using recurrent neural networks (RNNs) on SMILES. In particular, an RNN called Long Short-Term Memory (LSTM)\cite{hochreiter1997long} was employed. The  neural networks were pretrained on a general corpus of molecules, and then either fine-tuned on a small number of known actives, or coupled to an external scoring function to drive the generation of molecules with desired properties, e.g.~activity against biological targets.\cite{segler_generating_2018}
Both papers pointed out the then lack of models for direct graph generation, which is considerably harder than generating sequences.
Generative models based on neural networks have since been further explored, with work on 
VAEs,
\cite{janz_learning_2018, blaschke_application_2018, kusner_grammar_2017, dai_syntax-directed_2018, harel_prototype-based_2018, simonovsky_graphvae:_2018, samanta_designing_2018, liu_constrained_2018, jin_junction_2018, kajino_molecular_2018, lim_molecular_2018} 
RNNs,
\cite{yuan_chemical_2017, bjerrum_molecular_2017, ertl_silico_2017, neil_exploring_2018, olivecrona_molecular_2017, gupta_generative_nodate, jaques_sequence_2016, popova_deep_2018,  guimaraes_objective-reinforced_2017,sanchez-lengeling_optimizing_2017, li_learning_2018, li_multi-objective_2018, li_designing_2018}
generative adversarial networks (GANs),
\cite{guimaraes_objective-reinforced_2017, sanchez-lengeling_optimizing_2017, putin_reinforced_2018, de_cao_molgan:_2018}
adversarial autoencoders (AAEs),
\cite{kadurin_cornucopia_2016, kadurin_drugan:_2017, putin_adversarial_2018, polykovskiy_entangled_2018}
and graph convolutional networks,
\cite{de_cao_molgan:_2018, you_graph_2018, li_multi-objective_2018,assouel2018defactor}
using SMILES strings or graphs to represent molecules.

\begin{figure}
    \centering
    \includegraphics[width=0.8\textwidth]{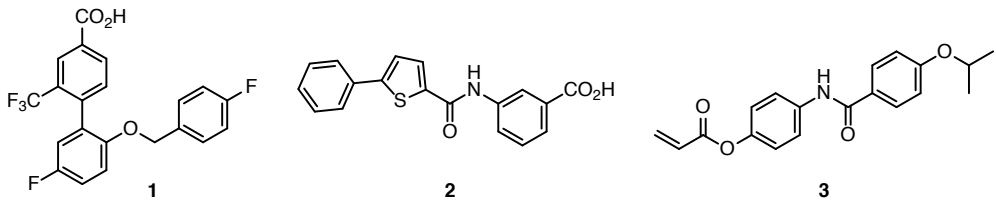}
    \caption{Compounds 1-3, generated by a SMILES-LSTM model,\cite{segler_generating_2018} were made and tested, and showed micro- to nanomolar activity against the RXR and PPAR receptors.\cite{merk2018novo,merk2018tuning}}
    \label{fig:recent_mols}
\end{figure}

Also, the first prospective studies using neural generative models for \denovo design have been published. Merk, Schneider and coworkers validated the SMILES-LSTM transfer learning method\cite{segler_generating_2018} by generating compounds \textbf{1-3} for nuclear receptors, which after synthesis and testing showed micro- to nanomolar activity (see Figure \ref{fig:recent_mols}).\cite{merk2018tuning,merk2018novo}

An important result about the space that pretrained neural models can explore was recently published by Arus-Pous et al.\cite{arus-pous_exploring_2018} They studied whether SMILES RNNs trained on a small subset (0.1\%) of GDB13, an enumerated molecule set, which covers all potentially stable molecules up to 13 heavy atoms, could recover almost the full dataset after sampling a sufficient number of compounds. They showed that the SMILES RNN was able to cover (and thus explore) a large portion of the complete chemical space defined by GDB13, and even generate molecules which were incorrectly omitted in the original GDB13.\cite{arus-pous_exploring_2018}

\section{Assessing De Novo Design Techniques}

To profile models for \denovo molecular design, we differentiate between their two main use cases:
\begin{itemize}
  \item Given a training set of molecules, a model generates new molecules following the same chemical distribution;
  \item A model generates the best possible molecules to satisfy a predefined goal.
\end{itemize}
The collection of benchmarks we propose below assesses both facets defined here. In the following we will refer to these two categories as \textit{distribution-learning} benchmarks and \textit{goal-directed} benchmarks, respectively.

The two benchmark categories are evaluated independently to afford models as much flexibility as possible without penalty, since there is no one-to-one correspondence between distribution-learning and goal-directed tasks. 
For instance, some models are able to generate optimized molecules without learning the chemical distribution first. Also, a model trained to reproduce a chemical distribution may then employ different strategies to deliver optimized molecules.

\subsection{Distribution-learning benchmarks}

Models for \denovo drug design often learn to reproduce the distribution of a training set, or use this training set to derive molecular fragments before generating targeted molecules. 
This allows some model architectures to learn the syntax of molecules in the selected molecular representation, and often accelerates the goal-directed tasks.

The distribution-learning benchmarks assess how well models learn to generate molecules similar to a training set, which in this work is a standardized subset of the ChEMBL database (see Appendix \ref{appendix:dataset}).\cite{mendez2018chembl} We consider five benchmarks for distribution learning:

\subsubsection{Validity}
The validity benchmark assesses whether the generated molecules are actually valid, i.e. whether they correspond to a (at least theoretically) realistic molecule. 
For example, molecules with an incorrect SMILES syntax, or with invalid valence, are penalized. 

\subsubsection{Uniqueness}
Given the high dimensionality of chemical space and the huge number of molecules potentially relevant in medicine, generative models should be able to generate a vast number of different molecules.
The uniqueness benchmark assesses whether models are able to generate unique molecules --- i.e., if a model generates the same molecule multiple times, it will be penalized.

\subsubsection{Novelty}

Since the ChEMBL training set represents only a tiny subset of drug-like chemical space, a model for \denovo molecular design with a good coverage of chemical space will rarely generate molecules present in the training set.
The novelty benchmark penalizes models when they generate molecules already present in the training set. Therefore, models overfitting the training set will obtain low scores on this task.

\subsubsection{Fréchet ChemNet Distance}
Preuer et al.~introduced the Fréchet ChemNet Distance (FCD) as a measure of how close distributions of generated are to the distribution of molecules in the training set.\cite{preuer_frechet_2018}
In the context of drug discovery, the usefulness of this metric stems from its ability to incorporate important chemical and biological features.
The FCD is determined from the hidden representation of molecules in a neural network called ChemNet trained for predicting biological activities, similarly to the Fréchet Inception Distance sometimes applied to generative models for images. More precisely, the means and covariances of the activations of the penultimate layer of ChemNet are calculated for the reference set and for the set of generated molecules. The FCD is then calculated as the Fréchet distance for both pairs of values. 
Similar molecule distributions are characterized by low FCD values.

\subsubsection{KL divergence}
The KL (Kullback-Leibler) divergence\cite{kullback1951information}
\begin{equation}
  D_\mathrm{KL}(P,Q) = \sum_i P(i) \log \frac{P(i)}{Q(i)}
\end{equation} 
measures how well a probability distribution $Q$ approximates another distribution $P$. 
For the KL divergence benchmark, the probability distributions of a variety of physicochemical descriptors for the training set and a set of generated molecules are compared, and the corresponding KL divergences are calculated.
Models able to capture the distributions of molecules in the training set will lead to small KL divergence values.

It has been noted that a lack of diversity of the generated molecules is an issue for a few models for \denovo design, in particular GANs.\cite{benhenda_chemgan_2017,preuer_frechet_2018} Other model classes do not suffer from that problem.\cite{segler_generating_2018} 
The KL divergence benchmark captures diversity to some extent, by requiring the generated molecules to be as diverse as the training set with respect to the considered property distributions.

\subsection{Goal-directed benchmarks}
\label{sec:goal_directed}
The goal-directed optimization of molecules relies on a formalism in which molecules can be scored individually. 
The molecule score reflects how well a molecule fulfills the required property profile (sometimes also called multi-property objective/MPO). 
The goal is to find molecules which maximize the scoring function. 

Concretely, the models are asked to generate a given number of molecules with high scores for a given function. The models have access to the scoring function and can iteratively improve their best molecule guesses, without knowing explicitly what the scoring function calculates. 

Here, by using robust and simple, but relevant scoring functions for molecular design, we disentangle the problem of molecule optimization from the problem of choosing good scoring functions, which has been highlighted many times in the context of \denovo design and virtual screening, and will not be the focus of this article.

In general, the function to optimize will be defined as the combination of one or several functions, representing different molecular features such as: 
\begin{itemize}
  \item Structural features. 
  Examples: molecular weight, number of aromatic rings, number of rotatable bonds;
  \item Physicochemical properties. 
  Examples: TPSA, logP;
  \item Similarity or dissimilarity to other molecules;
  \item Presence and absence of substructures, functional groups or atom types.
\end{itemize}

The majority of benchmarks presented in this work define complex combinations of such features.
In addition, some benchmarks fall into special categories:

\subsubsection{Similarity}
Similarity is one of the core concepts of chemoinformatics.\cite{willett1998chemical,gasteiger2006chemoinformatics} 
It serves multiple purposes and is an interesting objective for optimization. 
First, it is a surrogate for machine learning models, since it mimics an interpretable nearest neighbor model. 
However, it has the strong advantage over more complex machine learning (ML) algorithms that deficiencies in the ML models, stemming from training on small datasets or activity cliffs,  cannot be as easily exploited by the generative models. 
Second, it is directly related to virtual screening: \denovo design with a similarity objective can be interpreted as a form of inverse virtual screening, where molecules similar to a given target compound are generated on the fly instead of looking them up in a large database. 
In the similarity benchmarks, models aim to generate molecules similar to a target that was removed from the training set. 
Models perform well for the similarity benchmarks if they are able to generate many molecules that are closely related to a given target molecule.
Alternatively, the concept of similarity can be applied to exclude molecules that are too similar to other molecules.

\subsubsection{Rediscovery}
Rediscovery benchmarks are closely related to the similarity benchmarks described above.
The major difference is that the rediscovery task explicitly aims to rediscover the target molecule, not to generate many molecules similar to it. Rediscovery benchmarks have been studied in previous work.\cite{zaliani2009second, segler_generating_2018}

\subsubsection{Isomers}
For the isomer benchmarks, the task is to generate molecules that correspond to a target molecular formula (for example C$_7$H$_8$N$_2$O$_2$). 
The isomers for a given molecular formula can in principle be enumerated, but except for small molecules this number will in general be very large. 
The isomer benchmarks represent fully-determined tasks that assess the flexibility of the model to generate molecules following a simple pattern (which is \textit{a priori} unknown).
These benchmarks have no direct application in drug discovery.

\subsubsection{Median molecules}
In the median molecules benchmarks, the similarity to several molecules has to be maximized simultaneously; it is designed to be a conflicting task.\cite{brown2004graph} Besides measuring the obtained top score, it is instructive to study if the models also explore the chemical space between the target structures.

\subsection{Measuring Compound Quality}
\label{sec:mol_quality}

In previous works, several authors have highlighted that unrestricted \denovo design algorithms can generate compounds which are potentially unstable, reactive, laborious to synthesize, or simply unpleasant to the eye of medicinal chemists.\cite{ertl2012iade, gasteiger2007novo, hartenfeller2011enabling} Systems proposing too many unsuitable compounds lose trust, and will not be accepted by medicinal chemists. Unfortunately, explicitly encoding all the background knowledge medicinal chemists acquire with experience as an exhaustive list of unstable or undesirable substructures is challenging, if not impossible, due to the inherent subjectivity and context dependency --- e.g. a toxicity risk might be assessed differently in oncology and diabetology. QSAR models only cover this background knowledge in a limited way.

Nevertheless, a \denovo design benchmark would be incomplete without an assessment of compound quality. Here, following previous work, we employ rule sets which were compiled to decide whether compounds can be included into high-throughput screening (HTS) collections. It has to be noted though that these rules are controversial. Assuming that molecules filtered out by these patterns are all unsuitable, and that all compounds passing the filters are reasonable molecules, would be misguided. These filters are therefore taken as a high precision, low recall surrogate measure - we know that a filtered out molecule is probably bad, but we also know that our list of filters is incomplete.

We employ Walters’ \texttt{rd\_filters} implementation,\cite{rdfilters} using the SureChembl, Glaxo, PAINS (all retrieved from ChEMBL\cite{mendez2018chembl}), and in-house rule collections, to calculate the compound quality metrics. The rule collection is available in the Supporting Information.\cite{guacamol_package}

\section{Methods}

Our framework for benchmarking models for \denovo design, GuacaMol, is available as an open-source Python package and can be downloaded from the Internet at \url{www.github.com/BenevolentAI/guacamol}.\cite{guacamol_package}
GuacaMol provides user-friendly interfaces that allow researchers to couple their models to the benchmarking framework with minimal effort.

We selected five distribution-learning and twenty goal-directed benchmarks.
The exhaustive specifications and implementation details of the distribution-learning and goal-directed benchmarks are given in Appendices \ref{appendix:dist} and \ref{appendix:goal}, respectively.

For all chemoinformatics-related operations, such as handling or canonicalization\cite{schneider2015a} of SMILES strings, physicochemical property (logP, TPSA\cite{ertl2000a}) or similarity calculations, GuacaMol relies on the \texttt{RDKit} package.\cite{rdkit}

Several of the molecule generation approaches we studied require a training dataset. 
ChEMBL 24 was used for this purpose.\cite{mendez2018chembl}
The generation of the dataset is discussed and detailed in Appendix~\ref{appendix:dataset}.

In addition to the introduction of a benchmark suite for generative chemistry, this paper provides its evaluation over several baselines. The goal is to present a fair comparison of some recent methods that seem to achieve state of the art results in the field. These have been selected to represent a variety of methods, ranging from deep learning generative models to established genetic algorithms and more recent Monte Carlo Tree Search (MCTS).
Internal molecule representation was also taken into account, so that methods using SMILES strings were represented alongside others using a graph representation of atoms and bonds. For completeness and to provide a lower bound on the benchmark scores, two dummy baselines, performing random sampling and best of dataset-picking, are also provided.
All the baseline models assessed in this work are described in Appendix~\ref{appendix:baselines}.
Our implementation of those models is available on GitHub.\cite{guacamol_baselines_repository}

\section{Results and Discussion}

\subsection{Distribution-learning benchmarks}

\input{distribution_learning.tex}

Table~\ref{tab:results_distribution_learning} lists the results for the distribution-learning benchmarks. 
The random sampler model is a useful baseline for comparison because it shows what scores can be expected for good models on the KL divergence and Fréchet benchmarks. 
Unsurprisingly, its novelty score is zero since all the molecules it generates are present in the dataset.
The SMILES LSTM model sometimes produces invalid molecules. However, they are diverse, as attested by the uniqueness and novelty benchmarks, and closely resemble molecules from ChEMBL.
The Graph MCTS model is characterized by a high degree of valid, unique, and novel molecules. Nevertheless, it is not able to reproduce the property distributions of the underlying training set as precisely as the SMILES LSTM model.
The AAE model obtains excellent uniqueness and novelty scores. 
The molecules this model produces are, however, not as close to the dataset as the ones generated by the SMILES LSTM model.
ORGAN obtains poor scores for all the distribution-learning tasks.
More than half the molecules it generates are invalid, and they do not resemble the training set, as shown by the KL divergence and FCD scores. This might indicate mode collapse, which is a phenomenon often observed when training GANs.
The scores of the VAE model are not the best in any categories, but this model consistently produced relatively good scores for all the tasks.
Our findings indicate that simpler generative models (VAE, LSTM) are more powerful than more complex models. This is in accordance with similar results in the literature on text and molecule generation.\cite{tevet2018evaluating, liu_constrained_2018}

SMILES strings and their depictions for molecules generated by all six baseline models are available in the Supporting Information or on \url{https://benevolent.ai/guacamol}.

\subsection{Goal-directed benchmarks}

\input{goal_directed_results.tex}

\begin{figure}
    \centering
    \includegraphics[width=1.0\textwidth]{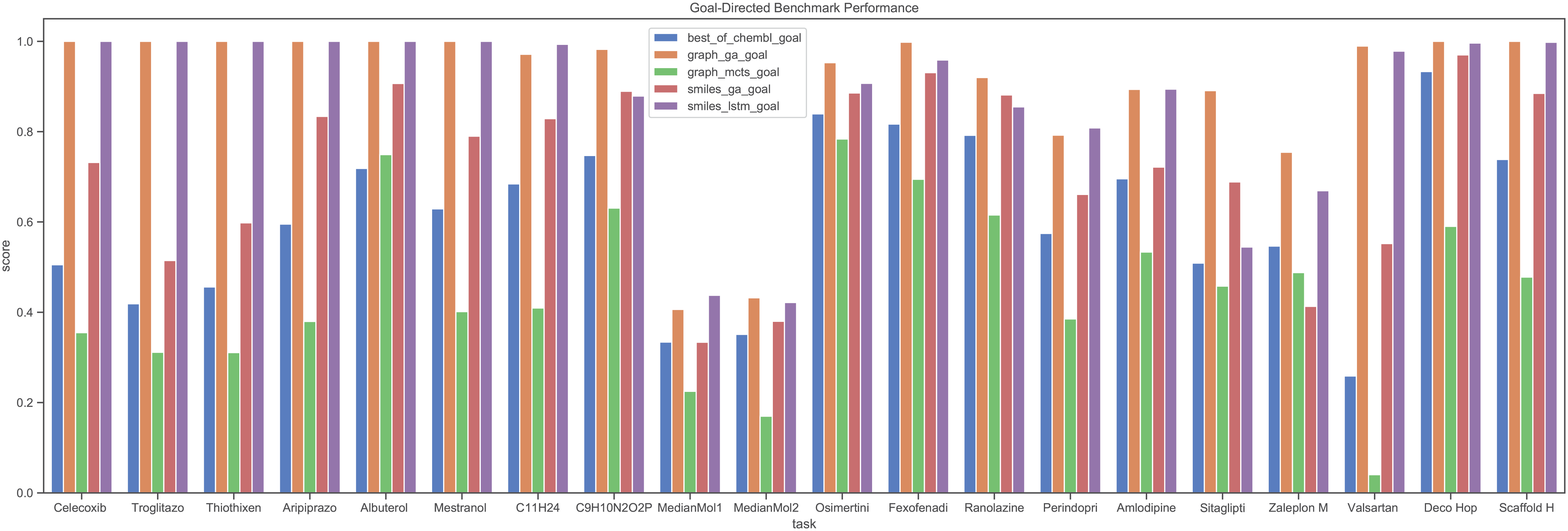}
    \caption{Results of the Goal-Directed Benchmarks.}
    \label{fig:results_goal_directed}
\end{figure}

Table~\ref{tab:results_goal_directed} and Figure~\ref{fig:results_goal_directed} show the results for the goal-directed benchmarks. 
The model ``Best of Dataset'' is a useful baseline, since it actually corresponds to a strategy based on virtual screening. It illustrates what minimal scores the models must obtain to have an advantage over pure virtual screening.
The Graph GA model obtains the best results for most benchmarks. 
The other GA model, based on SMILES strings, is distinctly better than the ``Best of Dataset'' baseline. However, especially for the similarity tasks, it scores lower than Graph GA.
The SMILES LSTM model nearly performs as well as the Graph GA model, and outperforms it on three benchmarks.
The Graph MCTS model performs worse than the baseline selecting the best candidates from ChEMBL.

In addition to these benchmarks, we explored a set of simpler benchmarks, mainly around physicochemical properties (e.g. logP and QED), which have been used as objectives in previous work on neural generative models (see Appendix~\ref{appendix:trivial}). However, we found that these benchmarks were not very well suited to differentiate between different models when appropriate baselines are employed. 

\begin{figure}
    \centering
    \includegraphics[width=0.7\textwidth]{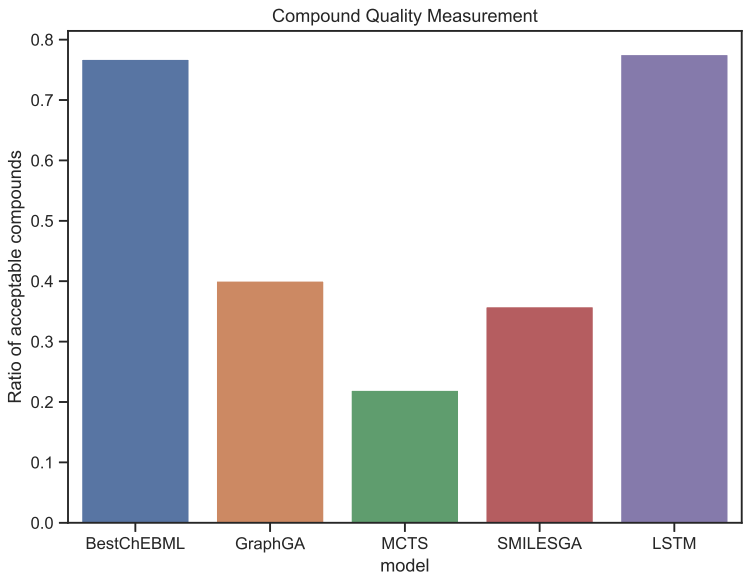}
    \caption{Quality of the molecules generated in the goal-directed benchmarks, assessed as the ratio of molecules passing the considered quality filters.}
    \label{fig:molquality}
\end{figure}

To measure the quality of the generated molecules from the goal-directed benchmarks, we gathered and analyzed the top 100 molecules from each task, and checked whether they passed the quality filters described earlier (see Figure~\ref{fig:molquality}). We found that 77\% of molecules from the Best of ChEMBL baseline pass the filters. Also, the same ratio is achieved by the LSTM model. The Graph GA, SMILES GA, and Graph MCTS produces 40\%, 36\%, and 22\% high quality molecules, respectively. This higher performance of the neural network based model could be explained by a transfer of the information the LSTM has learned about the general properties of the molecules’ distribution to the goal-directed optimization.

Overall, Graph MCTS was found to perform poorly, qualitatively and quantitatively, in the distribution-learning and goal-directed benchmarks. This could indicate that the search paradigm of starting with a single molecule at the root the search tree which is then subsequently refined in a search tree, might not be as well suited for \denovo design as approaches based on populations of good compounds. 
In concordance with previous work, the Graph GA performs extraordinarily well in optimization, due to its flexibility to make very fine grained local edits to the compounds, based on a population. However, again consistent with previous work, the GAs provide molecules of inconsistent quality, as it has no inherent or pre-encoded knowledge about a molecular distribution. 

The LSTM-based model performs slightly worse in the goal-directed benchmarks than the Graph GA, however the compound quality is considerably higher, being on the same level as the virtual screening baseline. This can be attributed to the transfer of learned information from the pretraining to the specific optimization problems. Neural models therefore seem to at least partially resolve the weaknesses of virtual screening and GAs, exhibiting both good optimization performance and molecule quality. 

SMILES strings and their depictions for molecules generated by the five baseline models for each MPO are available in the Supporting Information or on \url{https://benevolent.ai/guacamol}.

\section{Conclusions and Outlook}

Recently, generative models for \denovo molecular design based on neural networks have been introduced, and have shown promising results. However, there has been a lack of consistency in the evaluation of such models. The present work addressed this problem by introducing GuacaMol, a framework to quantitatively benchmark models for \denovo design. It aims to provide a standardized way of assessing new models, and to improve comparability of the models.

Our framework defines two evaluation dimensions. 
First, it assesses models for their ability to learn from a dataset of molecules and to generate novel molecules with similar properties.
Second, it evaluates the faculty of generative models to deliver molecules with specific properties, which we encode as molecule scores.

We provided a suite of benchmarks compatible with the evaluation framework.
It encompasses a wide range of tasks, designed to assess the flexibility of generative models.
The proposed suite comprises 5 benchmarks for the ability of generative models to learn distributions of molecules, 20 optimization benchmarks for the generation of molecules with specific properties, and a metric for compound quality. 

We evaluated a series of diverse baseline models.
In terms of optimization performance, the best model is a genetic algorithm based on a graph representation of molecules.
The second best optimizer is a recurrent neural network model that considers SMILES representations of molecules.
Both models achieve similar scores, which indicates that deep learning models can reach the optimization performance of gold-standard discrete algorithms for \denovo molecular design. However, when pre-trained on large datasets, the neural model outperforms the genetic algorithm in regards to compound quality.

The evaluation of the proposed benchmark suite for the baseline models pointed out that some benchmark tasks can be too easily solved by most of the models.
This indicates the necessity of harder tasks for benchmarking models for \denovo molecular design in the future. 

An important aspect that will need further focus is the quality of the generated molecules, which is difficult to measure objectively. Furthermore, depending on the requirements, time constraints or sample efficiency of different models can become important, and will require further attention. Also, more work is needed to move neural models from SMILES to graph representations, even though SMILES have been unreasonably effective.\cite{liu_constrained_2018} We are confident that the flexible open source framework presented herein will enable and inspire the community to come up with even more challenging benchmarks for \denovo design, so that researchers can rigorously assess their models, and computational chemists can confidently harness models with well-understood strengths and limitations.

\section{Acknowledgments}

The authors thank Nadine Schneider, Mohamed Ahmed, Matthew Sellwood, Joshua Meyers, Mark Rackham, Patrick Riley, and Amir Saffari for helpful discussions, and Kristina Preuer and Günter Klambauer for help with the FCD code.

\section{Appendix}

\subsection{Dataset Generation}
\label{appendix:dataset}

The datasets used for training generative models are derived from the ChEMBL 24 database.\cite{mendez2018chembl}
The advantage of ChEMBL is that it contains only molecules that have been synthesized and tested against biological targets. 
Other datasets, such as ZINC,\cite{irwin2005zinc} contain, at least to some degree, virtual molecules that are likely to be synthesizable, but have not been made yet. 
Furthermore, ZINC is biased towards smaller and more readily synthesizable molecules, indicated by a larger proportion of molecules containing amide bonds compared to ChEMBL. 
The QM9 set,\cite{ramakrishnan2014quantum} which is a subset of GDB9,\cite{ruddigkeit2012enumeration} is a completely enumerated dataset, which has shown to be of value in many applications. 
However, since it contains mostly compounds which have not been made yet, including many molecules with complex annulated ring systems, it is not very well suited to learn representations of drug-like, synthesizable molecules. 

To generate the final dataset for the benchmarks, ChEMBL is post-processed by 
\begin{enumerate}
  \item removal of salts;
  \item charge neutralization;
  \item removal of molecules with SMILES strings longer than 100 characters;
  \item removal of molecules containing any element other than H, B, C, N, O, F, Si, P, S, Cl, Se, Br, and I;
  \item removal of molecules with a larger\cite{rdkitblog} ECFP4 similarity than 0.323 compared to a holdout set consisting of 10 marketed drugs (celecoxib, aripiprazole, cobimetinib, osimertinib, troglitazone, ranolazine, thiothixene, albuterol, fexofenadine, mestranol).
  This allows us to define similarity benchmarks for targets that are not part of the training set.
\end{enumerate}

The post-processed ChEMBL dataset can be downloaded from the Internet.\cite{figshare} 
Additionally, a docker container with version-controlled dependencies to allow for reproducible dataset creation is provided in the \texttt{guacamol} repository.\cite{guacamol_package}

\subsection{Implementation details: Distribution-learning benchmarks}
\label{appendix:dist}

\paragraph{Validity}
The validity score is calculated as the ratio of valid molecules out of 10'000 generated molecules. 
Molecules are considered to be valid if their SMILES representation can be successfully parsed by \texttt{RDKit}.

\paragraph{Uniqueness}
To calculate the uniqueness score, molecules are sampled from the generative model until 10'000 valid molecules have been obtained. 
Duplicate molecules are identified by identical canonical SMILES strings.
The score is obtained as the number of different canonical SMILES strings divided by 10'000.

\paragraph{Novelty}
The novelty score is calculated by generating molecules until 10'000 different canonical SMILES strings are obtained, and computing the ratio of molecules not present in the ChEMBL dataset.

\paragraph{Fréchet ChemNet Distance (FCD)}
To generate the FCD score, a random subset of 10’000 molecules is selected from the ChEMBL dataset, and the generative model is sampled until 10'000 valid molecules are obtained. The FCD between both sets of molecules is calculated with the \texttt{FCD} package available on GitHub,\cite{fcd_package} and the final score $S$ is given by
\begin{equation}
    S = \exp (-0.2 \cdot \mathrm{FCD})
\end{equation}

\paragraph{KL divergence}
For this task, the following physicochemical descriptors are calculated with \texttt{RDKit} for both the sampled and the reference set:
\texttt{BertzCT}, \texttt{MolLogP}, \texttt{MolWt}, \texttt{TPSA}, \texttt{NumHAcceptors}, \texttt{NumHDonors}, \texttt{NumRotatableBonds}, \texttt{NumAliphaticRings}, and  \texttt{NumAromaticRings}.
Furthermore, we calculate the distribution of maximum nearest neighbour similarities on ECFP4 fingerprints for both sets. 
Then, the distribution of these descriptors is computed via kernel density estimation (using the \texttt{scipy} package) for continuous descriptors, or as a histogram for discrete descriptors. 
The KL divergence $D_{\mathrm{KL},i}$ is then computed for each descriptor $i$, and is aggregated to a final score $S$ via 
\begin{equation}
  S = \frac{1}{k}\sum_i^k \exp(-D_{\mathrm{KL},i}),
\end{equation}
where $k$ is the number of descriptors (in our case $k=9$).

\subsection{Implementation details: Goal-directed benchmarks}
\label{appendix:goal}

\subsubsection{Formalism}

In the goal-directed benchmarks discussed in this paper, raw molecular properties rarely correspond to the molecule scores used for optimization. Instead, they are post-processed by modifier functions that give more flexibility in how the final molecule score is computed, and furthermore restrict scores to the interval [0, 1].
For the benchmarks discussed below, we apply five different types of modifier functions. They are detailed in Appendix~\ref{appendix:modifiers}.

For many benchmarks the final molecule score corresponds to an average of multiple contributions. For instance, the molecule score for the median molecules benchmarks is an average of two similarity scores (see below). 
Depending on the benchmark, the average is calculated as the arithmetic mean or the geometric mean of the contributions.
With a geometric mean, all requirements (as given by the individual scoring functions) must be met, at least partly.
Otherwise, if one of the contributions has a score of zero, the final molecule score will also be zero.
With an arithmetic mean, the requirements are less strict, and a molecule can still achieve a good score if one of its partial scores is zero.

The final benchmark score is calculated as a weighted average of the molecule scores. For most of the benchmarks discussed below, the molecules with the best scores are given a larger weight. This choice reflects the compromise that models should be able to deliver a few molecules with top scores, but that they should also be able to generate many molecules with satisfactory scores. For all the goal-directed benchmarks, we calculate one or several average score for given numbers of top molecules, and then set the benchmark score to be the mean of these average scores. 
For instance, many benchmarks consider the combination of top-1, top-10 and top-100 scores, in which the benchmark score $S$ is given by 
\begin{equation}
    S = \frac{1}{3} \left(  s_1 + \frac{1}{10} \sum_{i=1}^{10} s_i + \frac{1}{100} \sum_{i=1}^{100} s_i \right)
\end{equation}
where $\boldsymbol{s}$ is a 100-dimensional vector of molecule scores $s_i$, $1 \leq i \leq 100$, sorted in decreasing order (i.e., $s_i \geq s_j$ for $i < j$).

The suite of goal-directed benchmarks considered in this work is given in Appendix~\ref{appendix:benchmark_list}.

During evaluation, the duration of the benchmark, the number of calls to the scoring function, and the internal similarity of the top molecules are captured in the results of each goal-directed benchmark, in order to allow for more fine-grained analyses.

\subsubsection{List of benchmarks}
\label{appendix:benchmark_list}

All the goal-directed benchmarks are listed in Table~\ref{tab:goal_directed_benchmarks}.
More information about the scoring functions and modifiers can be found in Appendices \ref{appendix:scoring_functions} and \ref{appendix:modifiers}, respectively.

\input{benchmark_list.tex}

The first six benchmarks fall into the rediscovery and similarity categories described in Section~\ref{sec:goal_directed}.
The molecules for these benchmarks are present in the holdout set considered during dataset generation.
The benchmarks C$_{11}$H$_{24}$ and C$_9$H$_{10}$N$_2$O$_2$PF$_2$Cl are isomer benchmarks.
Note that the molecule C$_{11}$H$_{24}$ has 159 possible isomers when ignoring stereochemistry. 
Therefore, for this benchmark the final score is obtained from the top-159 molecule scores.
The two following benchmarks are median molecule benchmarks that aim to generate molecules mixing camphor and menthol, as well as tadalafil and sildenafil.
The five following benchmarks (Osimertinib MPO, Fexofenadine MPO, Ranolazine MPO, Perindopril MPO, Amlodipine MPO), define objectives related to known drug molecules and trying to fine-tune their structural or physicochemical properties.
The Ranolazine MPO benchmark uses a start population, comprising one single molecule (ranolazine).
For the Sitagliptin MPO benchmark, the models must generate molecules that are as dissimilar to sitagliptin as possible, while keeping some of its properties.
In the Zaleplon MPO benchmark, generative models must find molecules that are similar to zaleplon but have a different molecular formula.
The valsartan SMARTS benchmark targets molecules containing a SMARTS pattern related to valsartan while being characterized by physicochemical properties corresponding to the sitagliptin molecule.
Finally, the Scaffold Hop and Decorator Hop benchmarks aim to maximize the similarity to a SMILES strings, while keeping or excluding specific SMARTS patterns, mimicking the tasks of changing the scaffold of a compound while keeping specific substituents, and keeping a scaffold fixed while changing the substitution pattern.

\subsubsection{Scoring functions}
\label{appendix:scoring_functions}

Most of the scoring functions used in this work can readily be calculated from SMILES strings or their underlying molecular graphs.
The scoring functions for TPSA, logP, Bertz and QED are calculated using the models contained in \texttt{RDKit}.
The similarity, SMARTS, and isomer scoring functions are explained in more detail in the following.

\paragraph{Similarity scoring function}

The similarity scoring function (called ``sim($x$, $y$)'' above) refers to a scoring function calculating the similarity to a target molecule $x$.
The score corresponds to the Tanimoto similarity when considering the fingerprint $y$ and is computed with \texttt{RDKit}.

\paragraph{SMARTS scoring function}

The SMARTS scoring function (called ``SMARTS($s$, $b$)'' above) evaluates whether the SMARTS pattern $s$ is present in the molecule to score or not. 
$b$ is a boolean value indicating whether the SMARTS pattern should be present (\textit{true}) or absent (\textit{false}). 
When the pattern is desired, a score of $1$ is returned if the SMARTS pattern is found, else $0$.
When the pattern should be absent, the score will be $0$ if the pattern is found, else $1$.
The pattern recognition is computed with \texttt{RDKit}.

\paragraph{Isomer scoring function}

For the isomer benchmarks, the molecule score (called ``isomer($x$)'' above) is calculated as the geometric mean of the following contributions:
\begin{itemize}
  \item For each element type present in the target molecular formula: 
  the number of atoms of this element type, modified by a Gaussian modifier.
  The Gaussian modifier has a mean corresponding to the target number of atoms of this element type, and a standard deviation of 1.0.
  \item The total number of atoms in the molecule, modified by a Gaussian modifier. 
  The Gaussian modifier has a mean corresponding to the total number of atoms of the target molecular formula, and a standard deviation of 2.0.
\end{itemize}
For instance, for the target formula C$_{11}$H$_{24}$, the score $s$ of a molecule $m$ is given by:
\begin{equation}
s(m) = \left( 
  \exp \left\{ -\frac{(n_\mathrm{C}(m) - 11)^2}{2} \right\} \cdot 
  \exp \left\{ -\frac{(n_\mathrm{H}(m) - 24)^2}{2} \right\} \cdot 
  \exp \left\{ -\frac{(n_\mathrm{tot}(m) - 35)^2}{8} \right\}
\right) ^{1/3}
\end{equation}
where $n_\mathrm{C}(m)$ and $n_\mathrm{H}(m)$ refer to the number of carbon and hydrogen atoms present in $m$, respectively, and $n_\mathrm{tot}(m)$ is the total number of atoms in $m$.

\subsubsection{Score modifiers}
\label{appendix:modifiers}
For the benchmarks discussed above, we apply different types of modifier functions:

\begin{itemize}
  \item \texttt{Gaussian}$(\mu, \sigma)$: 
  The Gaussian modifier allows to target a specific value of some property, while giving high scores when the underlying value is close to the target. It can be adjusted with the mean $\mu$ and standard deviation $\sigma$ of the underlying Gaussian function.

  \item \texttt{MinGaussian}$(\mu, \sigma)$: 
  The min Gaussian modifier corresponds to the right half of a Gaussian function. Values smaller than $\mu$ are given full score, and values larger than $\mu$ decrease continuously to zero.

  \item \texttt{MaxGaussian}$(\mu, \sigma)$: 
  The max Gaussian modifier corresponds to the left half of a Gaussian function. Values larger than $\mu$ are given full score, and values smaller than $\mu$ decrease continuously to zero.

  \item \texttt{Thresholded}$(t)$: 
  With the thresholded modifier, full score is attributed to values above a given threshold $t$. Values smaller than $t$ decrease linearly to zero.
\end{itemize}

\begin{figure}[htb]
    \centering
    \includegraphics[width=0.8\textwidth]{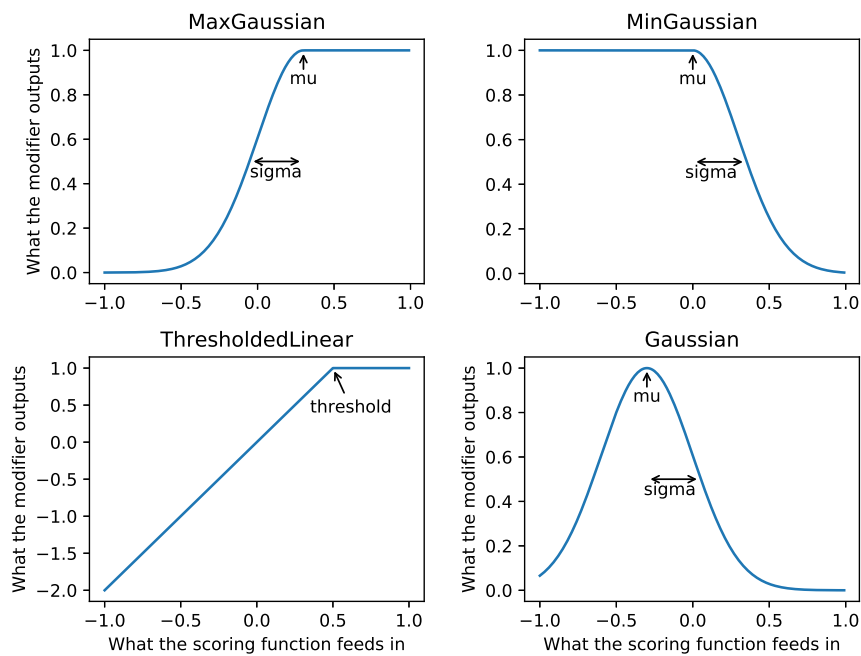}
    \caption{Modifier functions used in this study.}
    \label{fig:modifiers}
\end{figure}

The effect of these four modifier functions is illustrated in Figure~\ref{fig:modifiers}.

\subsection{Baseline Models}
\label{appendix:baselines}

Below, we give short descriptions of the baseline models involved in this work.
Our implementations of the baseline models detailed here are available on GitHub.\cite{guacamol_baselines_repository}

\subsubsection{Random sampler}
This baseline simply samples the requested number of molecules from the dataset at random. 
It provides a lower bound for the goal directed benchmarks as no optimization is performed to obtain the returned molecules, as well as an upper bound for two of the distribution learning benchmarks, as the molecules returned are taken directly from the original distribution.

\subsubsection{Best in dataset}
The goal of \denovo molecular design is to explore unknown parts of chemical space, generating new compounds with better properties than the ones already known. This baseline simply scores the entire dataset with the provided scoring function and returns the highest scoring molecules. This effectively provides a lower bound for the goal directed benchmarks as any good \denovo method should be able to generate better molecules than the ones provided during training or as a starting population. Distribution-learning benchmarks are not applicable to this baseline.

\subsubsection{SMILES GA}
Genetic algorithms (GA) for \denovo molecular design are well established technique.\cite{ hartenfeller2011enabling,brown2004graph}
Yoshikawa et al.~proposed a method that evolves string molecular representations using mutations exploiting the SMILES context-free grammar \cite{yoshikawa2018population}.

For each goal-directed benchmark the 300 highest scoring molecules in the dataset are selected as the initial population. Each molecule is represented by 300 genes. During each epoch an offspring of 600 new molecules are generated by randomly mutating the population molecules. After deduplication and scoring these new molecules are merged with the current population and a new generation is selected by keeping the top scoring molecules overall. This process is repeated 1000 times or until progress has stopped for 5 consecutive epochs. 
Distribution-learning benchmarks do not apply to this baseline.

\subsubsection{Graph GA}
The second genetic algorithm baseline follows the implementation of Jensen \cite{Jensen2018} in which molecule evolution happens at the graph level.

For each goal-directed benchmark the 100 highest scoring molecules in the dataset are selected as the initial population. During each epoch a mating pool of 200 molecules is sampled with replacement from the population, using scores as weights. This pool may contain many repeated specimens if their score is high. A new population of 100 is then generated by iteratively choosing two molecules at random from the mating pool and applying a crossover operation. With probability of 0.5 a mutation is also applied to the offspring molecule. This process is repeated 1000 times or until progress has stopped for 5 consecutive epochs. Distribution-learning benchmarks do not apply to this baseline.

\subsubsection{Graph MCTS}
The MCTS molecule generation procedure follows the implementation by Jensen.\cite{Jensen2018} The statistics used during sampling are computed on the GuacaMol dataset.

For this baseline no initial population is selected for the goal-directed benchmarks. Following the author's parameters, each new molecule is generated by running 40 simulations, starting from a CC base molecule, at each step 25 children are considered and the roll-out stops when reaching 60 atoms. The best scoring molecule found during the roll-out is returned. A population of 100 molecules is generated at each epoch. This process is repeated 1000 times or until progress has stopped for 5 consecutive epochs.

For the distribution-learning benchmark the generation starts from a CC base molecule and a new molecule is generated with the same parameters as for the goal oriented, the only difference being that no scoring function is provided, so the first molecule to reach terminal state is returned instead.

\subsubsection{SMILES LSTM}
The SMILES LSTM is a simple, yet powerful baseline model, consisting of a long-short term memory (LSTM)\cite{hochreiter1997long} neural network which predicts the next character of partial SMILES strings.\cite{segler_generating_2018} LSTM are more powerful than GRU networks, as they are able to learn a counting mechanism.\cite{weiss2018practical} Combined with a simple hill-climb algorithm for optimization, which is an off-policy policy gradient algorithm with binary rewards and can also be interpreted as iterative fine-tuning, LSTM has recently been shown to perform as well as more sophisticated reinforcement learning algorithms such as proximal policy optimization (PPO) or advantage actor critic (A2C).\cite{neil_exploring_2018}

The model used was an LSTM with 3 layers of hidden size of 1024. For the goal-directed benchmarks 20 iterations of hill-climbing were performed, at each step the model generated 8192 molecules and the top scoring 1024 were used to fine-tune the model parameters. For the distribution-learning benchmarks no fine tuning was done, the model simply generated the requested number of molecules.

\subsubsection{Variational Autoencoder}

Variational encoder\cite{gomez-bombarelli_automatic_2018} models learn a representation of molecules as latent vectors in a continuous space.
The model architecture comprises an encoder that converts SMILES strings to latent vectors, and a decoder that converts latent vectors back to SMILES strings.
A continuous-space representation of molecules enables sampling and property optimization in the latent space, after which the sample is decoded to produce the SMILES string of a molecule of interest.

We evaluate an implementation of VAE available on the Internet.\cite{moses_repository}
Only distribution-learning benchmarks are supported by this implementation. The VAE was trained with the hyperparameters previously optimized for the ZINC database. 

\subsubsection{Adversarial Autoencoder}

The approach taken by adversarial autoencoders\cite{polykovskiy_entangled_2018} is related the latent vector encoding of VAEs.
For AAEs, the latent space is regularized in a way to enforce a distribution in the latent space given by a prior.

We evaluate an implementation of AAE available on the Internet.\cite{moses_repository}
Only distribution-learning benchmarks are supported by this implementation. The AAE was trained with the hyperparameters previously optimized for the ZINC database.

\subsubsection{Objective-reinforced generative adversarial network}

The objective-reinforced generative adversarial network (ORGAN)\cite{guimaraes_objective-reinforced_2017} model architecture combines a generator and a discriminator network for molecule generation.

We evaluate an implementation of ORGAN available on the Internet.\cite{moses_repository}
Only distribution-learning benchmarks are supported by this implementation. The ORGAN was trained with the hyperparameters previously optimized for the ZINC database.

\subsection{Results for trivial goal-directed benchmarks}
\label{appendix:trivial}

Selecting adequate optimization objectives for benchmarking generative models is a challenging task.
The main difficulty is to find tasks that are hard enough, i.e.~objectives that are not already satisfied by molecules in the training dataset.
Otherwise, it is not possible to demonstrate the usefulness of generative models --- one must be able to assess whether (and to what extent) they are better than virtual screening approaches.

\input{trivial_list.tex}

Table~\ref{tab:trivial_benchmarks} lists seven easily-optimizable benchmarks, and Table~\ref{tab:results_trivial} shows the corresponding scores achieved by the baseline models.
The first three benchmarks in this table target given values for physicochemical properties.
All the models achieve near-perfect scores, including ``Best of dataset''.
The models also get perfect scores for the CNS MPO benchmark, which considers the CNS MPO score proposed by Pfizer.\cite{wager_central_2016}
The QED benchmark aims to optimize the ``quantitative estimate of drug-likeness'' (QED)\cite{bickerton_quantifying_2012}, an empirical measure of drug-likeness, similar to Lipinski's rule of 5.\cite{lipinski1997}
Even though optimization of drug-likeness alone is not a particularly useful objective in drug discovery, it has been used in several publications as a target for \denovo molecular design (see, for instance, Refs.~\citenum{sanchez-lengeling_optimizing_2017, gomez-bombarelli_automatic_2018, liu_constrained_2018, li_multi-objective_2018, you_graph_2018}).
The last two benchmarks, C$_7$H$_8$N$_2$O$_2$ and Pioglitazone MPO, involve more complex scoring functions but the generative models still achieve near-perfect scores for it.

\input{trivial_results.tex}

From these results, it is evident that such trivial benchmarks are not suitable for the assessment of generative models for \denovo molecular design.
The ChEMBL database alone achieves excellent scores already and therefore these benchmarks cannot demonstrate the advantage of generative models over pure virtual screening, and even less assess the difference in performance for the models.

\bibliography{denovo}

\end{document}

%% file: distribution_learning.tex
\begin{table}[hbt]
\caption{Results of the baseline models for the distribution-learning benchmarks}
\label{tab:results_distribution_learning}

\begin{adjustbox}{width=1\textwidth}
\begin{tabular}{|l|c|c|c|c|c|c|}

\hline
\textbf{Benchmark} & \textbf{Random sampler} & \textbf{SMILES LSTM} & \textbf{Graph MCTS} & \textbf{AAE} & \textbf{ORGAN} & \textbf{VAE} \\ 
\hline
Validity                                     & 1.000                                             & 0.959                                          & \textbf{1.000}                                & 0.822                                  & 0.379                                    & 0.870                         \\
Uniqueness                                   & 0.997                                             & \textbf{1.000}                                 & \textbf{1.000}                                & \textbf{1.000}                         & 0.841                                    & 0.999                         \\
Novelty                                      & 0.000                                             & 0.912                                          & 0.994                                         & \textbf{0.998}                         & 0.687                                    & 0.974                         \\
KL divergence                                & 0.998                                             & \textbf{0.991}                                 & 0.522                                         & 0.886                                  & 0.267                                    & 0.982                         \\
FCD                                          & 0.929                                             & \textbf{0.913}                                 & 0.015                                         & 0.529                                  & 0.000                                    & 0.863                         \\
\hline

\end{tabular}
\end{adjustbox}
\end{table}

%% file: goal_directed_results.tex
\begin{table}[hbt]
\caption{Results of the baseline models for the goal-directed benchmarks}
\label{tab:results_goal_directed}

\begin{adjustbox}{width=1\textwidth}
\begin{tabular}{|l|c|c|c|c|c|}

\hline
\textbf{Benchmark}              & \textbf{Best of dataset} & \textbf{SMILES LSTM} & \textbf{SMILES GA} & \textbf{Graph GA} & \textbf{Graph MCTS} \\
\hline
Celecoxib rediscovery                        & 0.505                                              & \textbf{1.000}                                 & 0.732                                        & \textbf{1.000}                              & 0.355                                         \\
Troglitazone rediscovery                     & 0.419                                              & \textbf{1.000}                                 & 0.515                                        & \textbf{1.000}                              & 0.311                                         \\
Thiothixene rediscovery                      & 0.456                                              & \textbf{1.000}                                 & 0.598                                        & \textbf{1.000}                              & 0.311                                         \\
Aripiprazole similarity                      & 0.595                                              & \textbf{1.000}                                 & 0.834                                        & \textbf{1.000}                              & 0.380                                         \\
Albuterol similarity                         & 0.719                                              & \textbf{1.000}                                 & 0.907                                        & \textbf{1.000}                              & 0.749                                         \\
Mestranol similarity                         & 0.629                                              & \textbf{1.000}                                 & 0.790                                        & \textbf{1.000}                              & 0.402                                         \\
C$_{11}$H$_{24}$                             & 0.684                                              & \textbf{0.993}                                 & 0.829                                        & 0.971                                       & 0.410                                         \\
C$_9$H$_{10}$N$_2$O$_2$PF$_2$Cl              & 0.747                                              & 0.879                                          & 0.889                                        & \textbf{0.982}                              & 0.631                                         \\
Median molecules 1                           & 0.334                                              & \textbf{0.438}                                 & 0.334                                        & 0.406                                       & 0.225                                         \\
Median molecules 2                           & 0.351                                              & 0.422                                          & 0.380                                        & \textbf{0.432}                              & 0.170                                         \\
Osimertinib MPO                              & 0.839                                              & 0.907                                          & 0.886                                        & \textbf{0.953}                              & 0.784                                         \\
Fexofenadine MPO                             & 0.817                                              & 0.959                                          & 0.931                                        & \textbf{0.998}                              & 0.695                                         \\
Ranolazine MPO                               & 0.792                                              & 0.855                                          & 0.881                                        & \textbf{0.920}                              & 0.616                                         \\
Perindopril MPO                              & 0.575                                              & \textbf{0.808}                                 & 0.661                                        & 0.792                                       & 0.385                                         \\
Amlodipine MPO                               & 0.696                                              & \textbf{0.894}                                 & 0.722                                        & \textbf{0.894}                              & 0.533                                         \\
Sitagliptin MPO                              & 0.509                                              & 0.545                                          & 0.689                                        & \textbf{0.891}                              & 0.458                                         \\
Zaleplon MPO                                 & 0.547                                              & 0.669                                          & 0.413                                        & \textbf{0.754}                              & 0.488                                         \\
Valsartan SMARTS                             & 0.259                                              & 0.978                                          & 0.552                                        & \textbf{0.990}                              & 0.040                                         \\
Deco Hop                                 & 0.933                                              & 0.996                                          & 0.970                                        & \textbf{1.000}                              & 0.590                                         \\
Scaffold Hop                                     & 0.738                                              & 0.998                                          & 0.885                                        & \textbf{1.000}                              & 0.478                                         \\ 
\hline
Total                                        & 12.144                                             & 17.340                                         & 14.396                                       & \textbf{17.983}                             & 9.009                                         \\ 
\hline

\end{tabular}
\end{adjustbox}
\end{table}

%% file: benchmark_list.tex
\begin{table}[p]
\caption{
  List of goal-directed benchmarks considered in this work.
  The molecule score given by each benchmark is a combination of one or several contributions, with their respective score modifiers.
  The column ``Scoring'' refers to the numbers of top molecules to consider in the score calculation.
  For clarity, the following SMARTS and SMILES strings were abbreviated in the table:
  $s_1 = \texttt{CN(C=O)Cc1ccc(c2ccccc2)cc1}$, 
  $s_2 = \texttt{[\#7]-c1n[c;h1]nc2[c;h1]c(-[\#8])[c;h0][c;h1]c12}$, 
  $s_3 = \texttt{[\#7]-c1ccc2ncsc2c1}$, 
  $s_4 = \texttt{CS([\#6])(=O)=O}$, 
  $s_5 = \texttt{CCCOc1cc2ncnc(Nc3ccc4ncsc4c3)c2cc1S(=O)(=O)C(C)(C)C}$, 
  and $s_6 = \texttt{[\#6]-[\#6]-[\#6]-[\#8]-[\#6]\newtilde[\#6]\newtilde[\#6]\newtilde[\#6]\newtilde[\#6]-[\#7]-c1ccc2ncsc2c1}$.
}
\label{tab:goal_directed_benchmarks}
\begin{adjustbox}{width=0.95\textwidth}
\begin{tabular}{|l|l|c|l|l|}

\hline
\textbf{Benchmark name}             & \textbf{Scoring}                        & \textbf{Mean}            & \textbf{Scoring function(s)}            & \textbf{Modifier}              \\ \hline
Celecoxxib rediscovery              & top-1                                   &                          & sim(Celecoxxib, ECFC4)                  & None                           \\ \hline
Troglitazone rediscovery            & top-1                                   &                          & sim(Troglitazone, ECFC4)                & None                           \\ \hline
Thiothixene rediscovery             & top-1                                   &                          & sim(Thiothixene, ECFC4)                 & None                           \\ \hline
Aripiprazole similarity             & top-1, top-10, top-100                  &                          & sim(Aripiprazole, ECFC4)                & \texttt{Thresholded}(0.75)     \\ \hline
Albuterol similarity                & top-1, top-10, top-100                  &                          & sim(Albuterol, FCFC4)                   & \texttt{Thresholded}(0.75)     \\ \hline
Mestranol similarity                & top-1, top-10, top-100                  &                          & sim(Mestranol, AP)                      & \texttt{Thresholded}(0.75)     \\ \hline
C$_{11}$H$_{24}$                    & top-159                                 &                          & isomer(C$_{11}$H$_{24}$)                & None                           \\ \hline
C$_9$H$_{10}$N$_2$O$_2$PF$_2$Cl     & top-250                                 &                          & isomer(C$_9$H$_{10}$N$_2$O$_2$PF$_2$Cl) & None                           \\ \hline
\multirow{2}{*}{Median molecules 1} & \multirow{2}{*}{top-1, top-10, top-100} & \multirow{2}{*}{geom.}   & sim(camphor, ECFC4)                     & None                           \\ \cline{4-5}
                                    &                                         &                          & sim(menthol, ECFC4)                     & None                           \\ \hline
\multirow{2}{*}{Median molecules 2} & \multirow{2}{*}{top-1, top-10, top-100} & \multirow{2}{*}{geom.}   & sim(tadalafil, ECFC6)                   & None                           \\ \cline{4-5}
                                    &                                         &                          & sim(sildenafil, ECFC6)                  & None                           \\ \hline
\multirow{4}{*}{Osimertinib MPO}    & \multirow{4}{*}{top-1, top-10, top-100} & \multirow{4}{*}{geom.}   & sim(osimertinib, FCFC4)                 & \texttt{Thresholded}(0.8)      \\ \cline{4-5}
                                    &                                         &                          & sim(osimertinib, ECFC6)                 & \texttt{MinGaussian}(0.85, 2)  \\ \cline{4-5}
                                    &                                         &                          & TPSA                                    & \texttt{MaxGaussian}(100, 2)   \\ \cline{4-5}
                                    &                                         &                          & logP                                    & \texttt{MinGaussian}(1, 2)     \\ \hline
\multirow{3}{*}{Fexofenadine MPO}   & \multirow{3}{*}{top-1, top-10, top-100} & \multirow{3}{*}{geom.}   & sim(fexofenadine, AP)                   & \texttt{Thresholded}(0.8)      \\ \cline{4-5}
                                    &                                         &                          & TPSA                                    & \texttt{MaxGaussian}(90, 2)    \\ \cline{4-5}
                                    &                                         &                          & logP                                    & \texttt{MinGaussian}(4, 2)     \\ \hline
\multirow{4}{*}{Ranolazine MPO}     & \multirow{4}{*}{top-1, top-10, top-100} & \multirow{4}{*}{geom.}   & sim(ranolazine, AP)                     & \texttt{Thresholded}(0.7)      \\ \cline{4-5}
                                    &                                         &                          & logP                                    & \texttt{MaxGaussian}(7, 1)     \\ \cline{4-5}
                                    &                                         &                          & TPSA                                    & \texttt{MaxGaussian}(95, 20)   \\ \cline{4-5}
                                    &                                         &                          & number of fluorine atoms                & \texttt{Gaussian}(1, 1)        \\ \hline
\multirow{2}{*}{Perindopril MPO}    & \multirow{2}{*}{top-1, top-10, top-100} & \multirow{2}{*}{geom.}   & sim(perindopril, ECFC4)                 & None                           \\ \cline{4-5}
                                    &                                         &                          & number aromatic rings                   & \texttt{Gaussian}(2, 0.5)      \\ \hline
\multirow{2}{*}{Amlodipine MPO}     & \multirow{2}{*}{top-1, top-10, top-100} & \multirow{2}{*}{geom.}   & sim(amlodipine, ECFC4)                  & None                           \\ \cline{4-5}
                                    &                                         &                          & number rings                            & \texttt{Gaussian}(3, 0.5)      \\ \hline
\multirow{4}{*}{Sitagliptin MPO}    & \multirow{4}{*}{top-1, top-10, top-100} & \multirow{4}{*}{geom.}   & sim(sitagliptin, ECFC4)                 & \texttt{Gaussian}(0, 0.1)      \\ \cline{4-5}
                                    &                                         &                          & logP                                    & \texttt{Gaussian}(2.0165, 0.2) \\ \cline{4-5}
                                    &                                         &                          & TPSA                                    & \texttt{Gaussian}(77.04, 5)    \\ \cline{4-5}
                                    &                                         &                          & isomer(C$_{16}$H$_{15}$F$_6$N$_5$O)     & None                           \\ \hline
\multirow{2}{*}{Zaleplon MPO}       & \multirow{2}{*}{top-1, top-10, top-100} & \multirow{2}{*}{geom.}   & sim(zaleplon, ECFC4)                    & None                           \\ \cline{4-5}
                                    &                                         &                          & isomer(C$_{19}$H$_{17}$N$_3$O$_2$)      & None                           \\ \hline
\multirow{4}{*}{Valsartan SMARTS}   & \multirow{4}{*}{top-1, top-10, top-100} & \multirow{4}{*}{geom.}   & SMARTS($s_1$, \textit{true})                           & None                           \\ \cline{4-5}
                                    &                                         &                          & logP                                    & \texttt{Gaussian}(2.0165, 0.2) \\ \cline{4-5}
                                    &                                         &                          & TPSA                                    & \texttt{Gaussian}(77.04, 5)    \\ \cline{4-5}
                                    &                                         &                          & Bertz                                   & \texttt{Gaussian}(896.38, 30)  \\ \hline
\multirow{4}{*}{Deco Hop}       & \multirow{4}{*}{top-1, top-10, top-100} & \multirow{4}{*}{arithm.} & SMARTS($s_2$, \textit{true})                           & None                           \\ \cline{4-5}
                                    &                                         &                          & SMARTS($s_3$, \textit{false})                  & None                           \\ \cline{4-5}
                                    &                                         &                          & SMARTS($s_4$, \textit{false})                  & None                           \\ \cline{4-5}
                                    &                                         &                          & sim($s_5$, PHCO)                        & \texttt{Thresholded}(0.85)     \\ \hline
\multirow{3}{*}{Scaffold Hop}      & \multirow{3}{*}{top-1, top-10, top-100} & \multirow{3}{*}{arithm.} & SMARTS($s_2$, \textit{false})                  & None                           \\ \cline{4-5}
                                    &                                         &                          & SMARTS($s_6$, \textit{true})                           & None                           \\ \cline{4-5}
                                    &                                         &                          & sim($s_5$, PHCO)                        & \texttt{Thresholded}(0.75)     \\ \hline
\end{tabular}
\end{adjustbox}
\end{table}

%% file: trivial_list.tex
\begin{table}[htb]
\caption{
  List of trivial optimization objectives in the goal-directed formalism. 
  Refer to Appendix~\ref{appendix:goal} for details about the formalism and implementation of goal-directed benchmarks.
}
\label{tab:trivial_benchmarks}
\begin{adjustbox}{width=1\textwidth}
\begin{tabular}{|l|l|c|l|l|}
\hline
\textbf{Benchmark name}           & \textbf{Scoring}                        & \textbf{Mean}      & \textbf{Scoring function(s)}   & \textbf{Modifier}                \\ \hline
logP (target: -1.0)               & top-1, top-10, top-100                  &                    & logP                           & \texttt{Gaussian}(-1, 2)         \\ \hline
logP (target: 8.0)                & top-1, top-10, top-100                  &                    & logP                           & \texttt{Gaussian}(8, 2)          \\ \hline
TPSA (target: 150.0)              & top-1, top-10, top-100                  &                    & TPSA                           & \texttt{Gaussian}(150, 2)        \\ \hline
\multirow{5}{*}{CNS MPO}          & \multirow{5}{*}{top-1, top-10, top-100} & \multirow{5}{*}{arithm.} & TPSA                           & \texttt{MinGaussian}(90, 2)      \\ \cline{4-5}
                                  &                                         &                    & TPSA                           & \texttt{MaxGaussian}(40, 2)      \\ \cline{4-5}
                                  &                                         &                    & number of hydrogen bond donors & \texttt{MinGaussian}(0, 2)       \\ \cline{4-5}
                                  &                                         &                    & logP                           & \texttt{MinGaussian}(5.0, 2)     \\ \cline{4-5}
                                  &                                         &                    & molecular weight               & \texttt{MinGaussian}(360, 2)     \\ \hline
QED                   & top-1, top-10, top-100                  &                    & QED                            & None                             \\ \hline
C$_7$H$_8$N$_2$O$_2$              & top-100                                 &                    & isomer(C$_7$H$_8$N$_2$O$_2$)   & None                             \\ \hline
\multirow{3}{*}{Pioglitazone MPO} & \multirow{3}{*}{top-1, top-10, top-100} & \multirow{3}{*}{geom.} & sim(pioglitazone, ECFP4)       & \texttt{Gaussian}(0, 0.1) \\ \cline{4-5}
                                  &                                         &                    & molecular weight               & \texttt{Gaussian}(356.447, 10)    \\ \cline{4-5}
                                  &                                         &                    & number rotatable bonds         & \texttt{Gaussian}(2, 0.5)        \\ \hline
\end{tabular}
\end{adjustbox}
\end{table}

%% file: trivial_results.tex
\begin{table}[htb]
\caption{Results of the baseline models for the trivial goal-directed benchmarks.}
\label{tab:results_trivial}
\begin{adjustbox}{width=1\textwidth}
\begin{tabular}{|l|c|c|c|c|c|}

\hline
\textbf{Benchmark}   & \textbf{Best of dataset} & \textbf{SMILES LSTM} & \textbf{SMILES GA} & \textbf{Graph GA} & \textbf{Graph MCTS} \\
\hline
logP (target: -1.0)  & 1.000                    & 1.000                & 1.000              & 1.000             & 0.986               \\
logP (target: 8.0)   & 1.000                    & 1.000                & 1.000              & 1.000             & 0.980               \\
TPSA (target: 150.0) & 1.000                    & 1.000                & 1.000              & 1.000             & 1.000               \\
CNS MPO              & 1.000                    & 1.000                & 1.000              & 1.000             & 1.000               \\
QED      & 0.948                    & 0.948                & 0.948              & 0.948             & 0.945               \\
C$_7$H$_8$N$_2$O$_2$ & 0.972                    & 1.000                & 0.992              & 0.993             & 0.851               \\
Pioglitazone MPO     & 0.982                    & 0.993                & 1.000              & 0.998             & 0.941               \\
\hline

\end{tabular}
\end{adjustbox}
\end{table}